\documentclass[runningheads]{llncs}

\usepackage[T1]{fontenc}
\usepackage{booktabs}
\usepackage{graphicx}
\usepackage{multirow}
\usepackage{amsfonts}
\usepackage{xcolor}

\usepackage{color}
\begin{document}
\title{DuDoUniNeXt: Dual-domain unified hybrid model for single and multi-contrast undersampled MRI reconstruction}
\titlerunning{DuDoUniNeXt}
%
\author{Ziqi Gao\inst{1,2}\and
Yue Zhang\inst{1,2}\and
Xinwen Liu\inst{3}\and
Kaiyan Li\inst{1,2}\and
S. Kevin Zhou\inst{1,2}}
\authorrunning{Gao et al.}
\institute{}
%
\institute{School of Biomedical Engineering, Division of Life Sciences and Medicine, University of Science and Technology of China, Hefei, Anhui, China 230026\and
Center for Medical Imaging, Robotics, Analytic Computing \& Learning (MIRACLE), Suzhou Institute for Advance Research, University of Science and Technology of China, Suzhou, Jiangsu, China 215123\and
School of Electrical Engineering and Computer Science, The University of Queensland}
%
\maketitle    
\begin{abstract}
Multi-contrast (MC) Magnetic Resonance Imaging (MRI) reconstruction aims to incorporate a reference image of auxiliary modality to guide the reconstruction process of the target modality. Known MC reconstruction methods perform well with a fully sampled reference image, but usually exhibit inferior performance, compared to single-contrast (SC) methods, when the reference image is missing or of low quality. To address this issue, we propose DuDoUniNeXt, a unified dual-domain MRI reconstruction network that can accommodate to scenarios involving absent, low-quality, and high-quality reference images. DuDoUniNeXt adopts a hybrid backbone that combines CNN and ViT, enabling specific adjustment of image domain and k-space reconstruction. Specifically, an adaptive coarse-to-fine feature fusion module (AdaC2F) is devised to dynamically process the information from reference images of varying qualities. Besides, a partially shared shallow feature extractor (PaSS) is proposed, which uses shared and distinct parameters to handle consistent and discrepancy information among contrasts. Experimental results demonstrate that the proposed model surpasses state-of-the-art SC and MC models significantly. Ablation studies show the effectiveness of the proposed hybrid backbone, AdaC2F, PaSS, and the dual-domain unified learning scheme.

\keywords{Unified model \and Undersampled MRI reconstruction  \and Multi-contrast MRI \and Hybrid model}
\end{abstract}
\section{Introduction}
Magnetic resonance imaging (MRI) is a superior imaging technique that provides multiple sequences of different modalities for comprehensive disease diagnosis. However, MRI typically involves long scanning time and expensive acquisition costs, which impose a heavy burden on patients. Mainstream methods for accelerating MRI can be categorized into \emph{single-contrast (SC) methods}~\cite{cheng2019spd,eo2018kiki,Guo2023ReconFormer,guo2021oucr,wang2016accelerating,yang2017dagan,quan2018compressedgan,schlemper2017dc,zhang2019reducing,Tolga2022unsupervised,huang2022sdaut,chung2022scoreMRI,MICCAI23huang2023CDiffMR,MICCAI23korkmaz2023selfsupervised,peng2022DiffuseRecon,adadiff2023} that reconstruct MRI from undersampled data of a single modality, and \emph{multi-contrast (MC) methods}~\cite{Tolga2022Ref,feng2022multi,liu2021deep,mimo-peng20a,zhou2020dudornet,multirecon2022Xuan,wang2017ismrm,mc2022herrmann} that utilize auxiliary easy-to-obtain modalities to assist in the reconstruction process of the target modality. Owing to the use of auxiliary information, MC methods usually achieve superior reconstruction quality.

\begin{figure}[t]
\centering
 
    \includegraphics[scale=0.5]{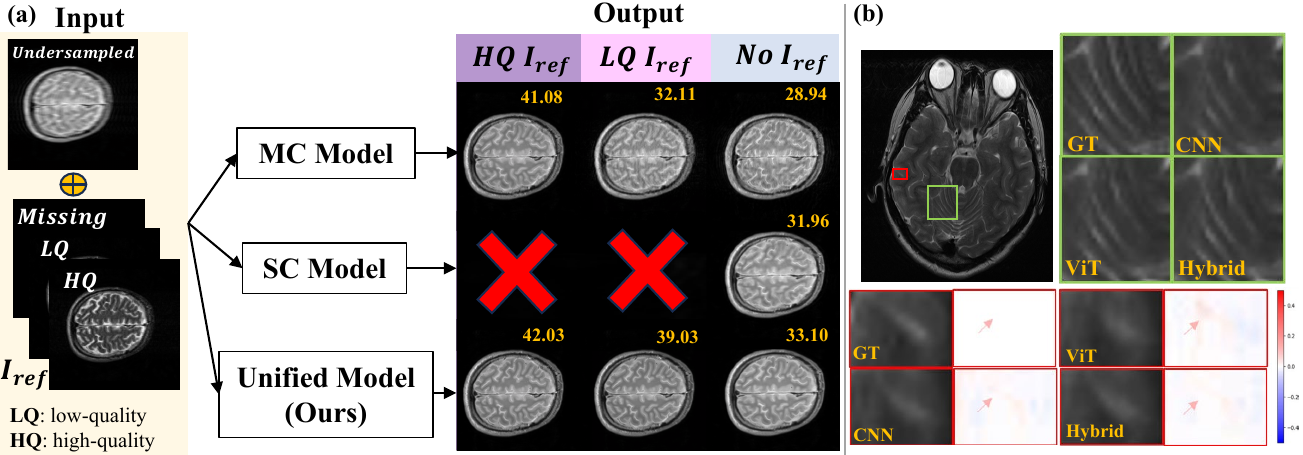}
    \caption{The main idea of DuDoUniNeXt. (a) The reconstruction results of different models with absent, an LQ, or an HQ auxiliary image. The upper part shows the results of the MC model, the middle part shows the results of the SC model, and the bottom is the results of the proposed DuDoUniNeXt. The PSNR values (dB) of the reconstructed images are shown in the upper right corners. 
    (b) The reconstruction emphasis of CNN, ViT, and our hybrid backbone. The \textcolor{green}{green} and \textcolor{red}{red} boxes highlight two typical patterns: highly structured lines and isolated fine details.
    }
    \vspace{-10pt} 
    \label{fig1_problem}
\end{figure}

Generally, MC reconstruction networks assume the reference image of auxiliary modality is fully sampled and of high quality \cite{liu2021regularization,lyu2022dudocaf,xiang2018ultra,zhou2023dsformer,zhou2020dudornet}. These networks reconstruct the undersampled images by effectively integrating complementary information from the reference image. RsGAN \cite{Tolga2022Ref} extends the adaptability towards a slightly downsampled (2-3$\times$) reference image. Although MC methods exhibit enhanced reconstruction quality, they inevitably fail in real-life scenarios where the reference images are missing or of low quality (e.g., undersampled).
For example, Fig.~\ref{fig1_problem} (a) shows the reconstruction results of SC model and MC model
with three types of reference images: missing, low-quality (LQ), or high-quality (HQ). 
Note that the MC model is MC-DuDoRNet from \cite{zhou2020dudornet} and the SC model is SC-DuDoRNet \cite{zhou2020dudornet}. It is observed that the MC model works well with an HQ auxiliary image (41.08dB in PSNR), but its performance degrades significantly with an LQ (32.11dB) or missing auxiliary image (28.94dB). In contrast, the SC model shows acceptable reconstruction performance (31.96 dB), however, it lacks the flexibility to incorporate HQ or LQ reference images for further performance enhancement. 

To solve the aforementioned issues, we propose a dual-domain unified dynamic
hybrid model, dubbed DuDoUniNeXt, for both single- and multi-contrast undersampled MRI reconstruction. Our DuDoUniNeXt can adaptively incorporate reference images of various qualities, and outperforms both MC and SC models (see Fig.~\ref{fig1_problem} (a)). The followings are our key contributions: 
\begin{itemize}
    \item \textbf{The first dual-domain unified network for both SC and MC MRI reconstruction.} DuDoUniNeXt is a novel dual-domain unified learning network that supports undersampled MRI reconstruction with high-quality, low-quality, or missing auxiliary images. The unified model has fewer parameters than the combination of two single-purpose models.
    \item \textbf{Contrast-aware dynamic encoder for MC MRI reconstruction.} Our encoder utilizes both complementary and consistent information from MC MRI concurrently with a Partially Shared Shallow feature extractor (PaSS) and supports dynamic MRI inputs with an Adaptive Coarse-to-Fine feature enhancement module (AdaC2F).
    \item \textbf{Dual-domain CNN-ViT hybrid backbone for MRI reconstruction.} The CNN-ViT hybrid backbone used by DuDoUniNeXt supports domain-specific model building combined with human heuristics, which surpasses unitary CNN or ViT backbones' efficacy with high efficiency (Fig.~\ref{fig1_problem} (b)).
    \item \textbf{Demonstrated robustness and effectiveness.} DuDoUniNeXt surpasses several popular single-contrast undersampled MRI reconstruction models on FastMRI and outperforms popular multi-contrast models on IXI considering multiple target-reference combinations.
\end{itemize}

\section{Method}
In this paper, we propose DuDoUniNeXt for unified SC and MC MRI reconstruction.
DuDoUniNeXt takes a dual-domain unified learning framework (Sec. \ref{sec:dudo}) consisting of recurrent k-space and image recovery networks. The reference image is injected into the image recovery network in each recurrent block with a contrast-aware dynamic encoder (Sec. \ref{sec:dynamic}). The backbone is an efficient CNN-ViT hybrid backbone that supports domain-specific adjustment (Sec. \ref{sec:hybrid}).

\begin{figure}[h]
    \centering
    \includegraphics[width=\textwidth]{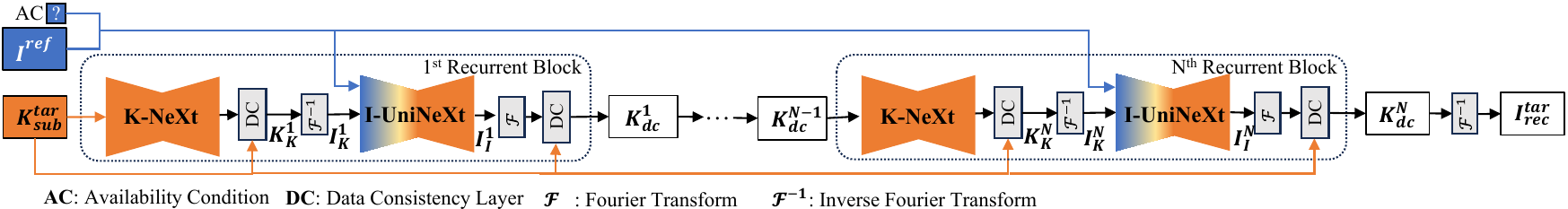}
    \caption{The overall dual-domain unified learning framework. 
    Each recurrent block contains one network for single-contrast k-space restoration, K-NeXt, and one network for unified image restoration, I-UniNeXt, with two interleaved DCs. Availability condition (AC) keeps I-UniNeXt notified for different conditions of $I_{ref}$.
    }
    \label{fig:dual-domain}
\end{figure}

\vspace{-10pt}\subsection{Dual-domain Unified Learning Framework}
\label{sec:dudo}
Given a missing, LQ, or HQ reference modality $I^{ref} \in \mathbf{C}^{hw}$,  a corresponding availability condition (AC) in \{0,1\} (with 0 denoting missing $I^{ref}$ and 1 denoting available $I^{ref}$), and undersampled k-space of the target modality $K^{tar}_{sub}\in \mathbf{C}^{hw}$,  DuDoUniNeXt aims to reconstruct the target image $I^{tar}_{rec}\in \mathbf{C}^{hw}$. 
As outlined in Fig.~\ref{fig:dual-domain}, DuDoUniNeXt uses recurrent learning with data consistency (DC) layers \cite{qin2018dc1,schlemper2017dc} for iterative feature recovery and adopts dual-domain learning \cite{zhou2020dudornet} to utilize the synergy of image and k-space domain learning.
The overall framework contains $N$ Recurrent Blocks, each of which performs one round of K-space (K-NeXt) and image-domain (I-UniNeXt) recovery on the input data.

Different from the existing dual-domain networks, DuDoUniNeXt uses a different image and k-space layout and reference modality injection approach to achieve unified learning. In the $i^{th} (i\in{1,...N})$  recurrent block, an estimation of the target modality is first given by a single-contrast k-space recovery network, K-NeXt. The input of K-NeXt is $K^{tar}_{sub}$ for the $1_{st}$ block and $K_{dc}^{i-1}$, the output of the previous recurrent block, for the others. K-NeXt's output that is enforced data consistency is $K_K^i$ and $I_{K}^{i}=\mathcal{F}^{-1}(K_K^i)$. $I_{K}^{i}$, together with the reference image $I_{ref}$ and AC, are sent to a unified multi-contrast image recovery network, I-UniNeXt. $K_{dc}^{i}$ is the output of I-UniNeXt after data consistency operation. The final output of DuDoUniNeXt is given by $I_{rec}^{tar}=\mathcal{F}^{-1}(K_{dc}^{N})$. 
The overall loss function is the summation of dual-domain $L_2$ loss of each recurrent block:
\vspace{-3pt}
\begin{equation}
    L = \sum_{i=1}^{N}(||K^{tar}_{GT}-K^{i}_K||_2+||I^{tar}_{GT}-\mathcal{F}^{-1}(K^{i}_{dc})||_2),
    \vspace{-2pt}
\end{equation}
in which $K^{tar}_{GT}$ and $I^{tar}_{GT}$ are the ground truth of the target K-space and image.

Intuitively, our DuDoUniNext can utilize the bias towards low-frequency in K-space learning with L$_2$ loss \cite{globalKspace2023} to give a rough estimation with K-NeXt and dynamically incorporate a reference to improve image details with I-UniNeXt.

\begin{figure}[t]
    \centering
    \includegraphics[width=\textwidth]{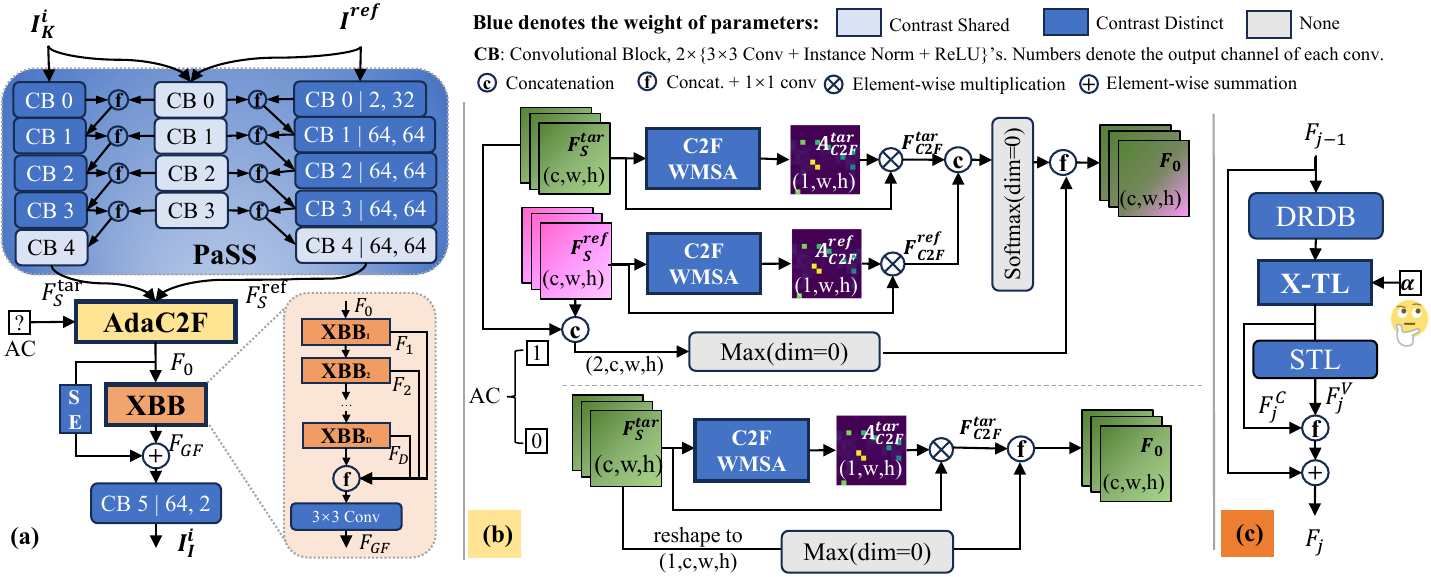}
    \caption{The architecture of I-UniNeXt. Panel (a) shows the overall architecture of I-UniNeXt, with the detailed architecture of the PaSS illustrated in the \textcolor{blue}{blue} box and XBB illustrated in the \textcolor{orange}{orange} box. (b) and (c) show the dedicated structure of AdaC2F and the domain-specific Hybrid Block (XBB$_j$), respectively.}
    \label{fig:i-uninext}
\end{figure}

\vspace{-10pt}\subsection{Contrast-aware Dynamic Encoder}
\label{sec:dynamic}
The key to the unified reconstruction model lies in how to leverage reference images of varying qualities, which is supported by the contrast-aware dynamic encoder in I-UniNeXt. Specifically, the encoder contains two core components: the Partially Shared Shallow feature extractor (PaSS) and Adaptive Coarse-to-Fine feature enhancement (AdaC2F).

\noindent {\bf Partially Shared Shallow Feature Extractor}
The target and reference images share the same anatomical structure and visualize soft tissue differently. 
To process both complementary and consistent information conveyed by multi-contrast inputs simultaneously, we propose PaSS to extract the multi-contrast shallow features with distinct and shared encoders, respectively. As depicted in Fig. \ref{fig:i-uninext}(a), PaSS consists of a shared branch and two modality-specific branches that have the same layer-wise convolution structure. For each modality, the consistent anatomical information captured by the shared branch is emphasized in the specific branches by layer-wise feature concatenation and 1x1 convolution. Two specific branches share the last convolution block (CB4) for memory efficiency and output modality-specific shallow features $F^{tar}_{s}$ and $F^{ref}_{s}$ respectively. The rationale behind it is the commonality of high-level information in MC MRI.

\noindent \textbf{Adaptive Coarse-to-fine Feature Enhancement}
The Max function is typically applied for multi-modal fusion \cite{fusion_max}, yet prone to detail loss and of limited ability to cope with the variance in the input's quality. To support both self and reference-enhancement for $F_S^{tar}$ and $F_S^{ref}$ adaptively, we propose AdaC2F that combines Max with multi-head self-attention (MSA) enhanced features reweighed by Softmax into a fused feature map $F_0$. We depict the variants of AdaC2F with different ACs in Figure \ref{fig:i-uninext}(b).
 The Softmax operates on the feature map enhanced by modality-specific coarse-to-fine windowed MSA (C2F-WMSA). Compared to the convolutional attention mechanism built on inductive bias, MSA is a data-specific operation \cite{park2022how}  with a large receptive field, enabling adaptively removing global artifacts. Considering that MSA has weaker detail preserving ability (Fig. \ref{fig1_problem}(b)), we compile two WMSAs with window sizes equal to $(16\times16)$ and $(8\times8)$ to produce modality-specific attention map $A^{*}_{C2F}$ that captures the global structure and emphasize the dedicated details:
\vspace{-4pt}
\begin{equation}
    F^{*}_{C2F} = F_S^{*} \odot A^{*}_{C2F}, A^{*}_{C2F} = \mathbf{WMSA}_{8}(\mathbf{WMSA}_{16}(F_S^{*})),
    \vspace{-4pt}
\end{equation}
where $\odot$ is the Hadamard product and * can be the target or reference branch.
\vspace{-6pt}\subsection{Domain-specific CNN-ViT hybrid backbone}
\label{sec:hybrid}
ViTs are more performant for undersampled MRI reconstruction than the CNN counterparts \cite{zhou2023dsformer,huang2022swinmr,lyu2022dudocaf}  at the cost of training difficulty and heavy computation. Also, ViT-based models typically have weaker detail preservation ability (Fig. \ref{fig1_problem}(b), \textcolor{red}{red} boxes). To balance between efficiency and efficacy, we propose a novel CNN-ViT hybrid backbone, XBB, consisting of D domain-specific hybrid blocks and a conventional global feature fusion operation as \cite{zhou2020dudornet}, shown in Fig. \ref{fig:i-uninext}(a), the \textcolor{orange}{orange} part. The design of each XBB is shown in Fig.~\ref{fig:i-uninext}(c). It is constructed by vertically stacking a Dilated Residual Dense Block (DRDB) \cite{gao2024dudornet+} and a Swin Transformer Layer (STL) \cite{liang2021swinir,huang2022swinmr,liu2021swin} followed by a $1\times1$ convolution operation that supports the hybrid interaction of CNN features $F_j^C$ and ViT features $F_j^V$. Intuitively, XBB utilizes the complementary properties of CNN (detail preservation) and ViT (feature aggregation) through this specific vertical layout design.

Another key property of XBB is that it can serve to build domain-specific models by adjusting the transition layer (X-TL).  X-TL is a $1\times1$ convolution that compresses $c$ feature maps into $\lfloor \alpha c \rfloor$, where $\alpha\in(0,1]$ is adjustable. Under a budget of total parameters, a larger $\alpha$ means more parameters for the ViT part and a smaller portion for CNN. This helps to build a domain-specific model by combining human heuristics: image restoration is considered a local problem \cite{liang2021swinir} so it may benefit from a smaller $\alpha_I$ while K-space shows a global dependency in recent K-space interpolation work \cite{globalKspace2023,gao2024dudornet+}; thus a larger $\alpha_K$ may help. 

 To further improve target contrast recovery, the globally fused feature $F_{GF}$ is combined with a global residual connection of a target-specific feature map enhanced by Squeeze-and-Excitation (SE) \cite{hu2018squeeze}. Feature is later passed through several convolutional layers, CB5, to generate the reconstruction output $I_I^i$:
\vspace{-3pt}
\begin{equation}
   I_I^{i} = \mathbf{SE}(F_0) + \mathbf{CB5}(F_{GF}), F_{GF}=\textbf{Conv}(\textbf{Concat}(F_1,...,F_D)).
   \vspace{-3pt}
\end{equation}

K-NeXt is a SC k-space recovery network consisting of XBB and CB5 only. 
\subsection{Training scheme} 
We employ a simple yet effective training strategy: exposing the network to various input configurations by providing HQ, LQ, and absent reference images randomly with equal probabilities of $(\frac{1}{3},\frac{1}{3},\frac{1}{3})$. Although we do witness an improvement by duplicating $I_{tar}^{sub}$ as input when $I_{ref}$ is missing, we use black images for simplified idea demonstration and controlled experiments.

\section{Experiment}
\subsubsection{Setup} In this section, we present the experimental setup including details of the dataset, implementation, and evaluation.

\noindent\underline{Dataset} We use one simulation dataset, the Multi-contrast IXI brain dataset\cite{ixi}, and one dataset containing raw coil-combined k-space, the FastMRI  knee\cite{fastmri1,fastmri2} dataset.
We use the well-aligned T$_2$- and PD-weighted coil-combined magnitude-only slices from 576 overlapped subjects in IXI and raw k-space coronal PD-weighted slices from 1,172 subjects with or without fat suppression in FastMRI. We follow the same pre-processing and dataset partition conventions as those in \cite{zhou2023dsformer} and \cite{chung2022scoreMRI,fastmri1}, respectively.  Our model, together with all other supervised models, is trained with 1D Cartesian sampling trajectories with a fixed portion (12.5\%) of the auto-calibration region, and the acceleration rates ranging from 4 to 8 $\times$ randomly and uniformly. 

\noindent\underline{Implementation}
 We implement our method in PyTorch using an NVIDIA GeForce RTX 3090 GPU with 24GB memory. The Adam solver \cite{kingma2014adam} is used to optimize DuDoUniNeXt with $lr = 0.0001, \beta_{1}\beta_{2}=0.5,0.999$. The training batch size is 1. I-UniNeXt and K-NeXt share the parameters among all recurrent blocks. In the main experiments, we simply use a moderate $\alpha_K=\alpha_I=0.5$; yet we do recognize performance gain by adjusting the rate (Table 2 in the supplementary).
 
 \noindent\underline{Evaluation} For comparision, we use the official implementations of D5C5~\cite{schlemper2017dc}, UNet~\cite{fastmri1}, DiffuseRecon~\cite{peng2022DiffuseRecon},
 Score-MRI~\cite{chung2022scoreMRI}, DuDoRNet~\cite{zhou2020dudornet}, and MTrans~\cite{feng2022multi}. We use the provided checkpoints trained on the same FastMRI dataset for the two diffusion models\cite{peng2022DiffuseRecon,chung2022scoreMRI} and train the rest models following their default settings unless otherwise noted. In Table 1 of the supplementary, we list a modified subset of hyper-parameters in DuDoRNet and MTrans to regulate the total parameters of single-purpose SoTA models. The final evaluation is performed on 5,145 slices of PD knee, 700 slices of T2w, and 700 slices of PD brain from unseen subjects during training. We use the Peak Signal-to-Noise Ratio (PSNR) and Structural Similarity Index (SSIM) to evaluate the reconstruction ability.

\begin{table*}[t]
\centering
    \resizebox{0.92\textwidth}{!}{
      \begin{tabular}{l|l|c|c|c|c|c|c}
        \toprule
        \multicolumn{2}{c|}{\textbf{PSNR(dB)/SSIM(\%)}} & \multicolumn{2}{c|}{\textbf{FastMRI}} & \multicolumn{4}{c}{\textbf{Multi-contrast IXI}} \\ 
        \midrule
        \multicolumn{2}{c|}{\textbf{Acc. rate}}  & 4x & 8x  & 4x & 8x  & 4x & 8x \\
        \midrule
            \multicolumn{2}{c|}{\textbf{SC reconstruction: target}}  & \multicolumn{2}{c}{\textbf{PD knee}}  & \multicolumn{2}{|c}{\textbf{T2w brain}} & \multicolumn{2}{|c}{\textbf{PD brain}} \\
        \midrule
        \multirow{6}{*}{SC Models} & \textbf{Zero Filling} 
        & 28.82/85.12  & 28.20/83.23  & 24.85/77.40  & 23.93/77.30  & 24.68/86.73  & 23.78/77.00     \\  
        & \textbf{UNet} & 30.71/88.99 & 29.18/85.63 & 27.76/87.97 & 24.54/85.38 & 27.67/87.59 & 24.13/85.27   \\
                  
        & \textbf{DiffuseRecon} 
        & 31.38/82.05
        & 29.51/76.23
        & N/A & N/A  & N/A & N/A\\
        & \textbf{Score-MRI} 
        & 30.33/87.39
        & 28.81/84.23
        & 28.78/93.69
        & 24.72/89.48
        & 28.64/87.59
        & 24.74/85.27\\
        & \textbf{D5C5} & 33.13/86.30 & 30.31/86.30 & 31.36/95.56 & 27.34/91.05  & 31.75/95.58  & 27.80/91.87\\    
        & \textbf{SC-DuDoRNet} 
        & 33.77/91.63
        & 31.12/87.78
        & 32.94/96.57
        & 28.50/92.75
        & 33.61/96.64
        & 29.25/93.51 \\
        & \textbf{SC-Dual Hybrid} & \textcolor{red}{33.91/91.78} & \textcolor{red}{31.18/87.91} & 33.65/97.01 & 28.67/93.07 & 34.26/97.04 & 29.57/93.89\\
        \midrule
        \multirow{4}{*}{MC Models} & \multicolumn{7}{l}{\textit{Without $I_{ref}$} }\\
        & \textbf{MTrans} & - & - & 27.16/86.34 & 25.92/85.92 & 27.34/86.13 & 26.01/85.50 \\
        & \textbf{MC-DuDoRNet} & - & - & 28.01/90.79 & 26.27/88.23 & 28.62/91.19 & 26.71/89.22\\
         & \textbf{Ours} & - & - & \textcolor{red}{34.14/97.27} & \textcolor{red}{28.91/93.46} & \textcolor{red}{34.56/97.30} & \textcolor{red}{29.62/94.23}\\
        \bottomrule
        \multicolumn{2}{c|}{\textbf{MC recon.: ref. \& tar.}}& \multicolumn{2}{c}{}& \multicolumn{2}{|c}{\textbf{ref:PD, tar:T2w}} & \multicolumn{2}{|c}{\textbf{ref:T2w, tar:PD}} \\
        \midrule
         \multirow{8}{*}{MC Models} & \multicolumn{7}{l}{\textit{With a LQ ($2\times$ under-sampled) $I_{ref}$}} \\
        & \textbf{MTrans} & - & - & 27.71/88.31 & 26.45/88.06 & 31.57/94.61 & 30.33/94.57 \\
         & \textbf{MC-DuDoRNet}  & - &- & 29.80/90.37 & 27.96/88.19 & 31.35/92.02 & 29.56/90.42\\
         & \textbf{Ours} & - & - & \textcolor{red}{37.97/98.73} & \textcolor{red}{35.66/98.21} & \textcolor{red}{38.74/98.83} &\textcolor{red}{36.38/98.35} \\
        \cmidrule{2-8}
        & \multicolumn{7}{l}{\textit{With a HQ (fully-sampled) $I_{ref}$}} \\
        & \textbf{MTrans} & - & - & 36.62/98.01 & 35.16/97.84 & 37.49/98.25 & 36.16/98.14 \\
        & \textbf{MC-DuDoRNet}  & - & - & 39.63/99.02 & 37.55/98.60 & 40.14/99.09 & 37.83/98.67\\
        & \textbf{Ours} & - & - & \textcolor{red}{40.22/99.12} &  \textcolor{red}{37.81/98.72} & \textcolor{red}{40.74/99.18} & \textcolor{red}{38.44/98.83} \\
        \bottomrule
      \end{tabular}
    }
    \caption{Quantitative comparisons on the reconstruction of brain and knee images under acceleration rates of $4\times$ and $8\times$ The \textcolor{red}{best} results are colored in \textcolor{red}{red}.}
    \label{main}
\end{table*}

\begin{figure}[t]
    \centering
    \includegraphics[width=\textwidth]{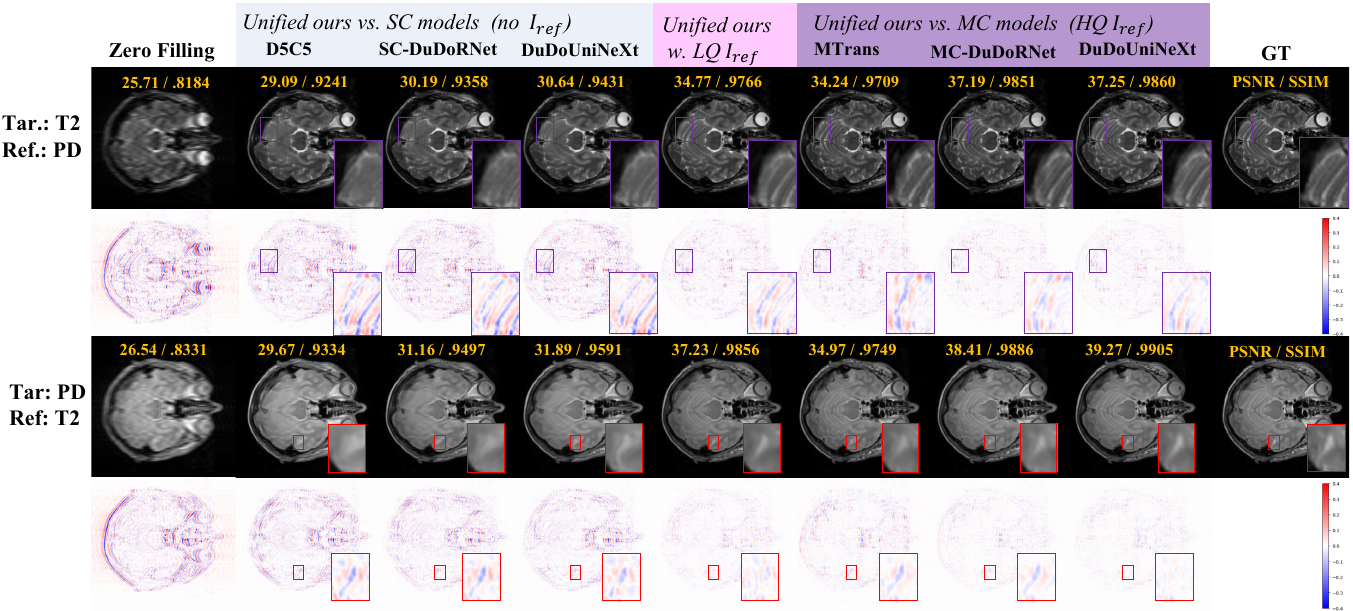}
    \caption{Qualitative comparisons of 5 $\times$ undersampled T2 and PD reconstruction under different $I_{ref}$ conditions. Corresponding error maps are illustrated in BWR colormaps. }
    \label{fig:vis}
\end{figure}

\vspace{-8pt}\subsubsection{Main results}
Quantitative evaluations on FastMRI and IXI are summarized in Table \ref{main}. The first sub-table validates the SC reconstruction ability of various methods on two datasets. On the one hand, by substituting the backbone of SC-DuDoRNet with our hybrid backbone, SC-Dual Hybrid improves the performance consistently for both datasets. On the other hand, our unified model, DuDoUniNeXt, surpasses the SC models with 0.4-1.2dB in PSNR. For comparison, other MC models have inferior performance when $I_{ref}$ is missing. For example, MC-DuDoRNet is 2.5-5dB lower than SC-DuDoRNet in terms of PSNR with the same input. The 2$^{nd}$ sub-table compares DuDoUniNeXt with some multi-contrast (MC) reconstruction methods by using paired (T2w, PD) from IXI to simulate two different combinations of target and reference. Adding an LQ $I_{ref}$, the reconstruction is boosted by 4-7dB in PSNR and further by another 2-3dB for an HQ $I_{ref}$. For comparison, MTrans and MC-DuDoRNet still lag behind SC-models with an LQ $I_{ref}$ as input and show a lower PSNR and SSIM than our DuDoUniNeXt under the specific fully-sampled reference-guided scenario.

Qualitative comparison of various T2 and PD reconstructions under different conditions of $I_{ref}$ is shown in Fig. \ref{fig:vis}. We also zoom in two typical regions: a highly structural region of the T2w image in \textcolor{violet}{violet} and an isolated small tissue of the PD image in \textcolor{red}{red}. Our model handles both structures and details better than existing SC and MC models.

\subsubsection{Ablation studies}
We validate the effectiveness of each novel component: AdaC2F, unified learning scheme, PaSS, XBB, and training scheme, and show the results in Table 2. (1) We demonstrate the superiority of AdaC2F by substituting it with existing feature fusion strategies: Max\cite{fusion_max}, HeMIS\cite{fusion_hemis}, DFUM\cite{zhang2023unified} and a reverse-version of ours, AdaF2C. (2) We verify our dual-domain learning schemes with different image and k-space network layouts and unification schemes. (3) We verify our PaSS by adjusting the $CB_i$'s with completely shared or completely distinct weights. (4) We substitute XBB with existing DRDN~\cite{zhou2020dudornet} and RSTB~\cite{liang2021swinir,huang2022swinmr,lyu2022dudocaf} as building blocks. More studies on domain-specificity and hybrid strategy are provided in the supplement. (5) We train existing MC models under our training scheme to show the training scheme is not the only reason for our performance gain. The superiority of DuDoUniNeXt's components is consistent.





\begin{table}[t]
\centering
\begin{minipage}[b]{0.5\textwidth}
\centering
\resizebox{\textwidth}{!}{%
\begin{tabular}{@{}l|l|ccc@{}}
\toprule
\multicolumn{2}{c|}{PSNR(dB)/SSIM(\%)} & HQ $I_{ref}$ & LQ $I_{ref}$ & Missing $I_{ref}$ \\ \midrule
\multirow{5}{*}{Fusion}& Max \cite{fusion_max} & 39.59/99.01 & 37.51/98.58 & 32.10/96.00\\
& HeMIS \cite{fusion_hemis} & 39.60/99.01 & 37.59/98.59 & 32.08/95.97 \\
& DFUM \cite{zhang2023unified} & 39.61/98.87 & 37.49/98.55 & 32.11/95.98\\
& AdaF2C & 39.33/98.96 & 37.46/98.62 & 32.22/95.88 \\
& AdaC2F (ours) & \textcolor{red}{39.63/99.03} & \textcolor{red}{37.68/98.63} & \textcolor{red}{32.39/96.19}\\
\midrule
Dual-domain & UniI-UniK & 38.37/98.69 & 36.79/98.21 & 30.91/94.68\\
Unified     & UniK-UniI & 39.24/98.96 & 37.63/98.61 & 31.85/96.79 \\
Learning & UniI-K & 39.09/98.83 & 37.61/98.50 & 31.36/95.14\\
& K-UniI (ours) & \textcolor{red}{39.63/99.03} & \textcolor{red}{37.68/98.63} & \textcolor{red}{32.39/96.19} \\
\bottomrule
\end{tabular}
}
\label{ab:fusion1}
\end{minipage}%
\hfill
\begin{minipage}[b]{0.49\textwidth}
\centering
\resizebox{\textwidth}{!}{%
\begin{tabular}{@{}l|l|ccc@{}}
\toprule
\multicolumn{2}{c|}{PSNR(dB)/SSIM(\%)} & HQ $I_{ref}$ & LQ $I_{ref}$ & Missing $I_{ref}$ \\ \midrule
Shallow& Shared & 39.13/98.93 & 37.18/98.50 & 32.04/95.94 \\
Feature & Distinct & 39.54/99.01 & 37.54/98.58 & 32.09/95.99\\
Extractor & PaSS (ours) & \textcolor{red}{39.63/99.03} & \textcolor{red}{37.68/98.63} & \textcolor{red}{32.39/96.19} \\
\midrule
Backbone& DRDN \cite{zhou2020dudornet} & 38.87/98.86 & 36.56/98.33 & 31.64/95.87 \\
Building & RSTB \cite{liang2021swinir,huang2022swinmr,lyu2022dudocaf} & 39.13/98.90 & 37.12/98.58 &31.95/96.01 \\
Block & XBB (ours) & \textcolor{red}{39.63/99.03} & \textcolor{red}{37.68/98.63} & \textcolor{red}{32.39/96.19}  \\
\midrule
Training & DuDoRNet w. TS & 39.01/98.89 & 33.93/95.12& 31.55/95.38 \\
Scheme  & MTrans w. TS & 35.05/97.50 & 33.03/96.40 & 28.16/91.04 \\
& Ours & \textcolor{red}{39.63/99.03} & \textcolor{red}{37.68/98.63} & \textcolor{red}{32.39/96.19} \\
\bottomrule
\end{tabular}
}
\label{ab:fusion2}
\end{minipage}%
\caption{Ablation studies. We use T2w images under different conditions as $I_{ref}$ to guide 5$\times$ undersampled PD reconstruction. The \textcolor{red}{best} results are colored \textcolor{red}{red}.}
\end{table}

\vspace{-5pt}\section{Conclusion}
We propose a novel approach for unified single-contrast and multi-contrast MRI reconstruction using a dual-domain network and an efficient CNN-ViT hybrid backbone. The experimental results on IXI and FastMRI demonstrate the superior adaptability and performance of DuDoUniNeXt against single-purpose SoTA models. The current study is based on the single-coiled images for a proof-of-concept and we will adapt it to multi-coil datasets in the future.

%
%
%
\newpage
\bibliographystyle{splncs04}
\bibliography{main}
\end{document}


\title{Supplementary materials of DuDoUniNeXt}
\author{Paper ID: 859}
\authorrunning{Paper ID: 859}
%
%
\titlerunning{}
\maketitle              
\section{Details of the architecture}

\begin{table}[]
\resizebox{\textwidth}{!}{
\begin{tabular}{l|l|l|l|l|l|l|l|l|l}
\toprule
                   &   Total Params.       & head\_dim    & emb\_dim & P1     & P2          \\
MC MTrans         & 2.1M          & 16$\to$4       & 256\to32   & 8\to4   & 16\to8         \\
\midrule
               & Total Params (two models)               & n\_recurrent & G0        & G      & D       & C       \\
SC/MC DuDoRNet  & 2*2.3M(I+K=1.1M+1.1M) & 5\to2        & 32\to64    & 32\to48 & 3\to4    & 4       \\
\midrule
                & Total Params.   & n\_recurrent & G0        & G      & D$_I$,D$_K$ & C$_I$,C$_K$ & \alpha$_I$, \alpha$_K$ & STL: window & STL: head\\
DuDoUniNeXt     & 3.6M (UniI+K=2.5M+1M)    & 2           & 64        & 48     & 4,3     & 5,3 & 0.5,0.5  & $16\times16$  & 4 \\   
\midrule
               &   Total Params. & n\_recurrent & G0        & G      & D       & C$_I$,C$_K$ &  \alpha$_I$, \alpha$_K$ & STL:window & STL: head \\
SC Dual Hybrid & 2.3M(I+K=1.1M+1.1M)& 2 & 64 & 32 & 4 & 4,4 &  0.5,0.5 & $16\times16$ & 4 \\
\bottomrule
\end{tabular}
}
\caption{Statistics of different model architectures. Others remain the same with their official implementations from Github. All models are trained for 100 epochs on IXI and 30 epochs on FastMRI and tested with the best checkpoints on validation sets.}
\label{hyperparameter}
\end{table}

\section{More ablation studies}
\noindent\textbf{XBB: domain-specific $\alpha_I$, $\alpha_K$}
\begin{table}[]
\centering
\begin{tabular}{l|l|l|l|l}
\toprule
PSNR(dB)/SSIM(\%) & \multicolumn{4}{c}{$\alpha$} \\
 & 0.25  & 0.5 & 0.75& 1 \\
\midrule
SC I-NeXt     &   31.15/95.62   &  \textcolor{red}{31.23/95.71} & 31.16/95.62  &  31.17/95.65  \\
SC K-NeXt     &   27.28/90.84   &  27.27/90.82  &  \textcolor{red}{27.28/90.85} & 27.28/90.83 \\
\bottomrule
\end{tabular}
\caption{Ablation study of XBB's domain-specific $\alpha$. SC I-NeXt and SC K-NeXt are single-domain models trained on the IXI PD datasets. PSNR and SSIM of the reconstruction of $5\times$ downsampled PD images of the testing set are shown. The \textcolor{red}{best} results are marked as \textcolor{red}{red}.}
\end{table}

\noindent\textbf{XBB: intra-stage v.s. inter-stage hybrid strategy.}
\\
\\
\\

\begin{figure}
    \centering
    \resizebox{\textwidth}{!}{
    \includegraphics{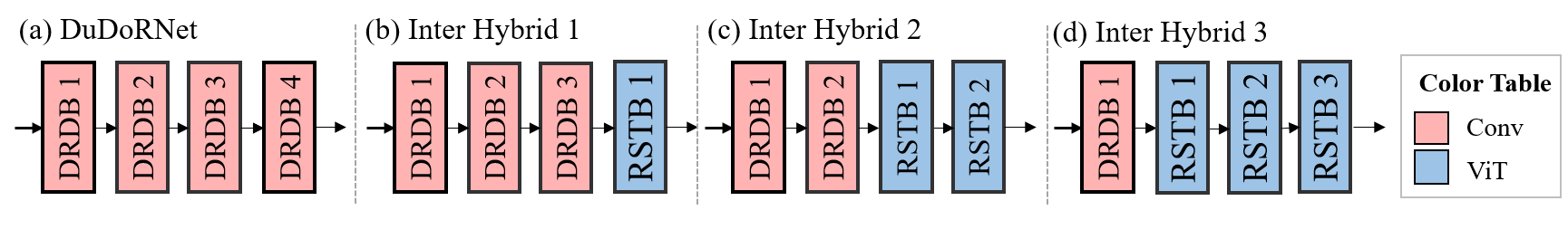}}
    \caption{The backbone of inter-stage models. The other parts stay the same with DuDoRNet's\cite{zhou2020dudornet}. (a) is the backbone of DuDoRNet, consisting of $D=4$ Dilated Residual Dense Blocks (DRDB)\cite{zhou2020dudornet}. (b)-(d) subtitutes DRDB from back to front with Residual Swin Transformer Blocks (RSTB) \cite{liang2021swinir,huang2022swinmr,lyu2022dudocaf}. }
    \label{fig:intra}
\end{figure}
\begin{figure}[t]
        \resizebox{\textwidth}{!}{
		\begin{tabular}{l|l|l|l|l|l}
        \toprule
           PSNR(dB)/SSIM(\%) & DuDoRNet  & IH 1 & IH 2 & IH 3  & XBB(Ours) \\
        \midrule
        Image domain model   &  30.70/95.25  & 30.74/95.27 & 30.78/95.29 &  31.17/95.66 & 31.32/95.71 \\
        K-space domain model   &  27.00/90.53  & 27.11/90.58 & 27.11/90.51  & 27.15/90.53 & 27.27/90.82 \\
        \midrule
        Inference time per slice & 4.8ms & 7.6ms  & 10.1ms & 13.4ms & 10.3ms \\
        \bottomrule
		\end{tabular}  
        }
      \label{hybrid}
      \caption {Ablation study of XBB's hybrid strategy. Quantitative comparison (PSNR, SSIM and Inference Time (ms)) of CNN model (DuDoRNet), inter-stage hybrid models, and our intra-stage hybrid model(XBB).}
\end{figure}

\bibliographystyle{splncs04}
\bibliography{main}